\documentstyle[12pt]{article}

\baselineskip=\normalbaselineskip

\ifx\TwoupWrites\UnDeFiNeD\else\target{\magstepminus1}{11.3in}{8.27in}
        \source{\magstep0}{7.5in}{11.69in}\fi
\newfont{\fourteencp}{cmcsc10 scaled\magstep2}
\newfont{\titlefont}{cmbx10 scaled\magstep3}
\newfont{\authorfont}{cmcsc10 scaled\magstep1}
\newfont{\fourteenmib}{cmmib10 scaled\magstep2}
        \skewchar\fourteenmib='177
\newfont{\elevenmib}{cmmib10 scaled\magstephalf}
        \skewchar\elevenmib='177
\makeatletter
\newcommand\nonsequentialeqnum{
        \@addtoreset{equation}{section}
        \def\theequation{\arabic{section}.\arabic{equation}}}
\newif\ifp@bblock  \p@bblocktrue
\newcommand\nopubblock{\p@bblockfalse}
\newcommand\topspace{\hrule height 0pt depth 0pt \vskip}
\newcommand\p@bblock{\begingroup \tabskip=\hsize minus \hsize
        \baselineskip=1.5\ht\strutbox \topspace-2\baselineskip
        \halign to\hsize{\strut ##\hfil\tabskip=0pt\crcr
        \the\Pubnum\crcr\the\date\crcr}\endgroup}

\renewcommand\titlepage{\ifx\TwoupWrites\UnDeFiNeD\null\vspace{-1.7cm}\fi
\vskip0.6cm
        \ifp@bblock\p@bblock \else\hrule height 0pt \relax \fi}
\makeatother
\newtoks\date
\newtoks\Pubnum
\newtoks\pubnum
\Pubnum={
YITP-96-57 \crcr
KUNS1418 \crcr
HE(TH)96/15 \crcr
hep-th/9610199
}
\newcommand{\frontpageskip}{\vspace{12pt plus .5fil minus 2pt}}
\renewcommand{\title}[1]{\frontpageskip
        \begin{center}{\titlefont #1}\end{center}\par}
\renewcommand{\author}[1]{\frontpageskip\par\begin{center}
        {\authorfont #1}\end{center}
        \nobreak
        }

\newcommand{\address}[1]{\par\begin{center}{\sl #1}\end{center}\par}

\renewcommand{\thanks}[1]{\footnote{#1}}
\renewcommand{\abstract}{\par\frontpageskip\centerline{
        \fourteencp Abstract}
        \vspace{8pt plus 3pt minus 3pt}}
\begin{document}

\titlepage

\renewcommand{\thefootnote}{\fnsymbol{footnote}}
\title{
Combinatorics of Solitons\\
\vskip5pt
in Noncritical String Theory
}

\author{
Masafumi Fukuma${}^{1\,}$\thanks{
e-mail address: {\tt fukuma@yukawa.kyoto-u.ac.jp}}
and Shigeaki Yahikozawa${}^{2\,}$\thanks{
e-mail address: {\tt yahiko@gauge.scphys.kyoto-u.ac.jp}}
}

\address{
${}^1$
Yukawa Institute for Theoretical Physics\\
Kyoto University, Kyoto 606-01, Japan \\
{}~\\
${}^2$
Department of Physics\\
Kyoto University, Kyoto 606-01, Japan \\
}

\newcommand{\bc}{\begin{center}}
\newcommand{\ec}{\end{center}}
\newcommand{\bl}{\begin{flushleft}}
\newcommand{\el}{\end{flushleft}}
\newcommand{\bi}{\begin{itemize}\begin{enumerate}}
\newcommand{\ei}{\end{enumerate}\end{itemize}}
\newcommand{\bt}{\begin{tabbing}}
\newcommand{\et}{\end{tabbing}}
\newcommand{\np}{\newpage}
\newcommand{\pset}{\setcounter{page}{1}}
\newcommand{\wh}[1]{w^{({#1})}}

\renewcommand{\thefootnote}{\arabic{footnote}}
\setcounter{footnote}{0}
\newcommand{\cleqn}{\setcounter{equation}{0} \indent}
\renewcommand{\theequation}{\arabic{equation}}
\newcommand{\beqa}{\begin{eqnarray}}
\newcommand{\eeqa}{\end{eqnarray}}
\newcommand{\n}{\nonumber}
\newcommand{\nn}{\nonumber \\ }
\newcommand{\eq}[1]{(\ref{#1})}
\newcommand{\bC}{{\bf C}}
\newcommand{\cD}{{\cal D}}
\newcommand{\cH}{{\cal H}}
\newcommand{\cO}{{\cal O}}
\newcommand{\Psid}{\Psi^\dagger}
\newcommand{\norm}[1]{{\parallel {#1} \parallel}^2}
\newcommand{\nnorm}[1]{{{\parallel {#1} \parallel}^{\prime\,2}_l}}
\newcommand{\del}{\partial}
\newcommand{\db}{{\bar{\delta}}}
\newcommand{\gbar}{{\bar{g}}}
\newcommand{\dl}
           {\left[\,\frac{dl}{\,l\,}\,\right]}
\newcommand{\Det}{\,\mbox{Det}\,}
\newcommand{\Tr}{\,\mbox{Tr}\,}
\newcommand{\ldot}{\dot{l}}
\newcommand{\const}{\mbox{const.\ }}
\newcommand{\bra}[1]{\left\langle\,{#1}\,\right|}
\newcommand{\ket}[1]{\left|\,{#1}\,\right\rangle}
\newcommand{\bracket}[2]{
	\left\langle\left.\,{#1}\,\right|\,{#2}\,\right\rangle}
\newcommand{\vev}[1]{\left\langle\,{#1}\,\right\rangle}
\newcommand{\bZ}{{\rm {\bf Z}}}
\newcommand{\svac}{\bra{\sigma}}
\newcommand{\pvac}{\ket{\Phi}}
\newcommand{\rphi}{\varphi}
\newcommand{\rphih}{\hat{\rphi}}
\newcommand{\dphi}{\del\rphi}
\newcommand{\dphih}{\del\hat{\rphi}}
\newcommand{\cb}{\bar{c}}
\newcommand{\gint}{\oint^{\,p}}
\newcommand{\dds}{\frac{d s}{2\pi i}}
\newcommand{\ddz}{\frac{d\zeta}{2\pi i}}

\begin{abstract}

We study the combinatorics of solitons in $D<2$ (or $c<1$)
string theory.
The weights in the summation over multi-solitons
are shown to be automatically determined
if we further require that the partition function with soliton
background be a $\tau$ function of the KP hierarchy,
in addition to the $W_{1+\infty}$ constraint. \\

\end{abstract}


One of the clues to understanding nonperturbative behavior of
strings is that nonperturbative effects in string theory generally
have the form $e^{-{\rm const\,}/g}$, where $g$ is the closed string
coupling constant \cite{s}.
In fact, these effects naturally appear in D-brane theory \cite{p1,p2-g}
and M-theory \cite{m},
and play important roles in their dynamics.
However, the origin of such dependence is still not understood well
because we do not have the self-contained closed string field theory
such that both elementary and solitonic excitations of strings
can be treated in a systematic manner.
Therefore, it should be of great importance
to investigate these nonperturbative effects
in simpler models such as $D<2$ string theory,
which can be explicitly constructed and also exactly solved
as the double scaling limits of matrix models \cite{bk-ds-gm}.

In our previous paper \cite{fy}, we explicitly constructed soliton
operators in the Schwinger-Dyson equation approach to
$D<2$ (or $c<1$) string theory,
and investigated the nonperturbative effects due to these solitons.
{}Furthermore, we suggested that fermions should be regarded as
the fundamental dynamical variables in (noncritical) string theory,
since both elementary strings (macroscopic loops) and solitons are
constructed in their bilinear forms.
The purpose of the present letter is to study the combinatorics of
solitons in noncritical string theory, namely,
how to determine the weights in the summation over multi-solitons.

We start our discussion with reinvestigating the celebrated string
equations \cite{bk-ds-gm}.
{}For pure gravity ($D=1$ (or $c=0$) string),
it is represented as the Painlev\'{e} I equation:
\beqa
  4\,u(t,g)^2\,+\,\frac{2g^2}{3}\,\del_t^{\,2}\,u(t,g)\,=\,t,
  \label{1}
\eeqa
where $u(t,g)$ is the connected two-point function\footnote{
In matrix models with even potentials,
there arises the ``doubling phenomenon''
and the connected two-point function would be identified
with $f(t,g)\equiv 2\,u(t,g)$.
}
of cosmological terms $\cO_1$, $u(t,g)=\vev{\cO_1 \cO_1}_c$,
and $t$ and $g$ are, respectively, the (renormalized) cosmological
and string coupling constant.
The solution of this equation has the following form of asymptotic
genus expansion:
\beqa
  u_{\rm pert}(t,g)\,=\,\sum_{h=0}^{\infty}
  	\,u_{\rm pert}^{(h)}(t) \,g^{2h}.
  \label{2}
\eeqa
The nonperturbative corrections \cite{bk-ds-gm,d,s}
to this asymptotic solution,
$\Delta u(t,g)\equiv u(t,g)-u_{\rm pert}(t,g)$, can be
evaluated by expanding the string equation around $u_{\rm pert}(t,g)$.
In fact, if we first make a linear approximation and also
take into account only the leading term in $g$,
then we obtain
$\Delta u\sim \sqrt{g}\,t^{-1/8}\,e^{-4\sqrt{6}\,t^{5/4}/\,5 g}$.
Then treating nonlinear terms in the string equation as perturbation,
we can calculate higher corrections as\footnote{
Note that $\sqrt{g}\,t^{-5/8}$ (and thus $t^{5/4}/g$) is a dimensionless
combination of $g$ and $t$.
}
\beqa
  \Delta u(t,g) \cong t^{1/2}\sum_{n=1}^\infty a_n\,
	\left(\sqrt{g}\,t^{-5/8}\right)^n\,
	e^{- n\,4\sqrt{6}\,t^{5/4}/\,5g} ,
  \label{3}
\eeqa
where the coefficients $a_n$ satisfy the recursion equation
$(n^2-1)\,a_n + \sum_{k=1}^{n-1} a_k\,a_{n-k}=0$, and are solved to be
\beqa
  a_n\,=\,\left(-\,\frac{1}{6}\right)^{n-1}\,n\,a^{n}.
  \label{4}
\eeqa
The first coefficient $a_1=a$ is one of the constants of
integration for the string equation\footnote{
Another one is fixed by demanding the asymptotic
form $u_{\rm pert}^{(0)}(t)=-\sqrt{t}/2$.
Here the minus sign is chosen
so that $\Delta u(t,g)$ takes a real value.}
and remains undetermined.
This parameter is analogous to the theta parameter in QCD
which cannot be seen in perturbation theory.
Thus, this simple analysis shows that, once 1-soliton
($n=1$ case in the above equation) exists, namely,
$a$ does not vanish,
then a series of multi-solitons also must exist and be
summed up with definite weights,
so that the solution satisfies the string equation.
In the generic unitary case, $(p,q)=(p,p+1)$,
the same conclusion holds except that
the exponential factors in eq.\ \eq{3} are generalized to the form
$\exp \left(-\,\vec{n}\cdot\vec{\alpha}\,\,t^{1+1/2p}/g\right)$,
where $\vec{n}$ resides on the dominant integral lattice\footnote{
Here we consider only stable solitons
with negative value of exponents.}
of some dimension (2 for $p=3$ and 4 for $p=4$, for example),
and the vector $\vec{\alpha}=(\alpha_1,\alpha_2,\cdots)$
consists of the exponents of 1-soliton ({\em i.e.} linearized)
solutions.

In order to systematically study the above result,
we here briefly review our formulation given in ref.\ \cite{fy}.
We first introduce complex $\zeta$ plane on which live
$p$ pairs of free fermions $c_a(\zeta),
{}~\cb_a(\zeta)~(a=0,1,\cdots,p-1)$.
We then construct $p$ bosons $\rphi_a(\zeta)$ through bosonization:
\beqa
  \dphi_a(\zeta)\,=\,:\cb_a(\zeta)c_a(\zeta):,
\eeqa
which in turn give the fermions as
\beqa
  \cb_a(\zeta)\,=\,K_a\,:e^{\rphi_a(\zeta)}:,~~~~~~~
  c_a(\zeta)\,=\,K_a\,:e^{-\rphi_a(\zeta)}:.
\eeqa
Here $K_a$ is the cocycle factor and ensures the correct anticommutation
relations between different indices $a\neq b$,
and all operators are normal ordered with respect to $SL(2,\bC)$
invariant vacuum $\bra{{\rm vac}}$.
The string field which describes elementary excitations of strings
is identified with the macroscopic operator $\Psi(l)$ \cite{bdss}
which creates a loop boundary of length $l$.
The Laplace transforms of their (generally disconnected) correlation
functions are represented in terms of the boson $\dphi_0(\zeta)$ as
\beqa
  \vev{\dphi_0(\zeta_1)\cdots\dphi_0(\zeta_N)}
  &\equiv&\int_0^\infty
	dl_1\cdots dl_N\,e^{-l_1\zeta_1-\cdots-l_N\zeta_N}\,
	\vev{\Psi(l_1)\cdots\Psi(l_N)} \n\\
  &=&\frac{
    	\left\langle\left.\left.\left.\,
	-\frac{B}{g}\,\right|\,
	:\dphi_0(\zeta_1)\cdots\dphi_0(\zeta_N):
	\,\right|\,\Phi\,\right\rangle\right.}{
    	\left\langle\left.\left.\,
	-\frac{B}{g}\,\right|\,
	\,\Phi\,\right\rangle\right.},
\eeqa
where the normal ordering again respects $\bra{{\rm vac}}$.
In the above expression, the state $\bra{-B/g}$ is defined by
\beqa
  \bra{-\frac{B}{g}}\,\equiv\,
	\left\langle\,\sigma\,\left|\,\exp\left\{-\frac{1}{g}
	\gint\ddz B(\zeta)\,\dphi_0(\zeta)\right.\right\}\right. ,
\eeqa
where $\bra{\sigma}$ is the vacuum with $\bZ_p$-twist operator
$\sigma(\zeta)$ inserted at the point of infinity,
$\bra{\sigma}=\bra{{\rm vac}}\,\sigma(\infty)$ \cite{dfms-br}.
$B(\zeta)$ is the background which characterizes the theory,
and for the minimal $(p,q)$ case it has the form
$B(\zeta)=\sum_{n=1}^{p+q}B_n\,\zeta^{n/p}$ with nonvanishing $B_{p+q}$.
The contour should surround $\zeta=\infty$ $p$ times,
since the bosons $\dphi_a(\zeta)$ have the monodromy as
$\svac \dphi_a(e^{2\pi i}\zeta)=\svac \dphi_{[a+1]}(\zeta)$
with $[a]\equiv a\,({\rm mod}~p)$.
{}Furthermore, $\dphi_0(\zeta)$ has the mode expansion under $\svac$
as
\beqa
  \svac \dphi_0(\zeta) \,=\,
	\frac{1}{p}\sum_{n\geq1}\zeta^{-n/p-1}\svac\alpha_n,
\eeqa
and thus $\bra{-B/g}$ can be rewritten as
$\svac \exp\left\{-(1/g)\sum_{n=1}^{p+q}B_n\,\alpha_n\right\}$.

The Schwinger-Dyson equations \cite{fkn1-dvv,gn,g,fkn2-fkn3} are compactly
expressed by the requirement
that the state $\pvac$ is a {\em decomposable} state
satisfying the $W_{1+\infty}$ constraint\footnote{
The equivalence between the Schwinger-Dyson equations and the Douglas
equations \cite{md} is proved in ref.\ \cite{fkn2-fkn3}.}
\beqa
  W^k_n\,\pvac=0~~~(k\geq1,~n\geq -k+1).
\eeqa
Here the generators of the $W_{1+\infty}$ algebra \cite{winf}
are given by the mode expansion
\beqa
  W^k(\zeta)\,\equiv\,\sum_{n\in\bZ}W^k_n\,\zeta^{-n-k}
  \,=\, \sum_{a=0}^{p-1}:\cb_a(\zeta)\,\del_\zeta^{k-1}\,c_a(\zeta):.
\eeqa
In general, a state $\pvac$ is called decomposable
if it is written as $\pvac=e^H \ket{\sigma}$,
where $H$ is a bilinear form of fermions
and $\ket{\sigma}\equiv \sigma(0)\ket{{\rm vac}}$.
This is also equivalent to the statement that
$\tau(x)=\svac \exp\{\sum_{n=1}^\infty x_n\alpha_n\} \pvac$ is
a $\tau$ function of the KP hierarchy \cite{djkm}.

Connected correlation functions of macroscopic loops
are obtained as the cumulants of the correlation functions above,
and have the following dependence on the string coupling constant $g$:
\beqa
  {}F[j(\zeta)]
  &\equiv&\vev{\exp\left\{
  	\gint\ddz j(\zeta)\dphi_0(\zeta)\right\}-1}_c \n\\
  &=&\log \frac{
    	\left\langle\left.\left.\left.\,
	-\frac{B}{g}\,\right|\,
	:\exp\left\{\gint\ddz j(\zeta)\dphi_0(\zeta)\right\}:
	\,\right|\,\Phi\,\right\rangle\right.}{
    	\left\langle\left.\left.\,
	-\frac{B}{g}\,\right|\,
	\,\Phi\,\right\rangle\right.}\n\\
  &=&\sum_{h=0}^\infty\sum_{N=1}^\infty\,\frac{g^{-2+2h+N}}{N!}\,
	\gint\frac{d\zeta_1}{2\pi i}\cdots\gint\frac{d\zeta_N}{2\pi i}
	\n\\
  &&~~~~~~~\times\,j(\zeta_1)\cdots j(\zeta_N)\,
	\vev{\dphi_0(\zeta_1)\cdots\dphi_0(\zeta_N)}^{(h)}_c\,.
\eeqa

Soliton backgrounds are constructed by inserting
the {\em soliton operators} \cite{fy}
\beqa
  D_{ab}\,=\,\gint\ddz\,S_{ab}(\zeta)~~~~(a\neq b)
\eeqa
into correlation functions,
where the soliton fields $S_{ab}(\zeta)$ are defined by
\beqa
  S_{ab}(\zeta)&\equiv&\cb_a(\zeta)\,c_b(\zeta)~~~(a\neq b)\n\\
  &=&\left\{
        \begin{array}{ll}
          -\,K_a K_b\,:e^{\rphi_a(\zeta)-\rphi_b(\zeta)}:~~~~&(a<b)\\
          +\,K_a K_b\,:e^{\rphi_a(\zeta)-\rphi_b(\zeta)}:~~~~&(a>b)\,.
        \end{array}
        \right.
\eeqa
As shown in ref.\ \cite{fy}, the soliton operators commute with
the generators of the $W_{1+\infty}$ algebra,
$\left[W^k_n,\,D_{ab}\right]=0$ $(k\geq1,~n\in\bZ)$,
and thus, if the state $\pvac$ satisfies the $W_{1+\infty}$ constraint,
then the multi-soliton state
$D_{a_1 b_1}D_{a_2 b_2}\cdots\pvac$ also satisfies the same constraint.
{}Furthermore the nonperturbative effect from 1-soliton background
was evaluated and found to have the form
$e^{-{{\rm const}}\,/g}$ \cite{fy}.
Nonperturbative effects for multi-soliton backgrounds are
then roughly estimated by adding the exponents and found to
construct the same lattice with the one described in the beginning of
the present paper.

At first sight these multi-solitons could be summed in a
completely arbitrary manner,
since each multi-soliton state satisfies the $W_{1+\infty}$ constraint.
On the other hand, our analysis of the string equation shows that
the weights in the summation are fixed
except for a few undetermined parameters
(see eqs.\ \eq{3} and \eq{4}).
This paradox can be solved if we further require that
the multi-soliton state be decomposable.
In fact, since $D_{ab}$ is a fermion bilinear,
the multi-soliton states must have the following form
if we impose this decomposability condition:
\beqa
  \ket{\Phi,\theta}\,=\,\prod_{a\neq b}e^{\,\theta_{ab}\,D_{ab}}\,\pvac,
\eeqa
where $\pvac$ is also a decomposable state
satisfying the $W_{1+\infty}$ constraint,
and $\theta_{ab}$ ($a\neq b$) are arbitrary constants.
Thus, we find that the generating function for macroscopic loops
with soliton backgrounds is generally given by
\beqa
  {}F_\theta[j(\zeta)]
  &\equiv&\vev{\exp\left\{
  	\gint\ddz j(\zeta)\dphi_0(\zeta)\right\}-1}_{c,\,\theta} \n\\
  &=&\log\,\frac{
    	\left\langle\left.\left.\left.\,
	-\frac{B}{g}\,\right|\,
	:\exp\left\{\gint\ddz j(\zeta)\dphi_0(\zeta)\right\}:
	\,\right|\,\Phi,\theta\,\right\rangle\right.}{
    	\left\langle\left.\left.\,
	-\frac{B}{g}\,\right|\,
	\,\Phi,\theta\,\right\rangle\right.}\,,
\eeqa
where the parameters $\theta_{ab}$ $(a\neq b)$ remain undetermined.

We here make a comment that the nonperturbative effects from
multi-solitons can be explicitly calculated
for connected correlation functions of lower dimensional operators
$\cO_1,\,\cO_2\,\cdots,\cO_{p+q}$.
In fact, these connected correlation functions are obtained
by taking derivatives with respect to the background sources as
\beqa
  &&\vev{\cO_1^{\,m_1}\cO_2^{\,m_2}\cdots
	\cO_{p+q}^{\,m_{p+q}}}_{c,\,\theta}\n\\
  &&~~~~~~=\,
  (-g)^{\,m_1+m_2+\cdots+m_{p+q}}\,
  \frac{\del^{\,m_1}}{\del B_1^{\,m_1}}
  \frac{\del^{\,m_2}}{\del B_2^{\,m_2}}
  \cdots\frac{\del^{\,m_{p+q}}}{\del B_{p+q}^{\,m_{p+q}}}\,
  \log \left\langle\left.\,-\frac{B}{g}\,
	\right|\,\Phi,\theta\,\right\rangle\,,
\eeqa
and $\log \left\langle\left.-B/g\,\right|\,\Phi,\theta\,
\right\rangle$ can be evaluated at $\theta=0$ by rewriting it as follows:
\beqa
  \log\left\langle\left.-\frac{B}{g}\,\right|\,\Phi,\theta\,
 	\right\rangle
  \,=\,\log\left\langle\left.-\frac{B}{g}\,\right|\,\Phi\,\right\rangle
  \,+\,\log\,\vev{\prod_{a\neq b}e^{\,\theta_{ab}\,D_{ab}}},
\eeqa
where
\beqa
  \vev{\prod_{a\neq b}e^{\,\theta_{ab}\,D_{ab}}}\,=\,
	\frac{
    	\left\langle\left.\left.\left.\,
	-\frac{B}{g}\,\right|\,
	\prod_{a\neq b}e^{\,\theta_{ab}\,D_{ab}}
	\,\right|\,\Phi\,\right\rangle\right.}{
    	\left\langle\left.\left.\,
	-\frac{B}{g}\,\right|\,
	\,\Phi\,\right\rangle\right.}\,.
\eeqa

As an example, we consider the $(p,q)=(2,3)$ case (pure gravity),
setting the background as
$B_1=t,\,B_{5}=-15/8$ and $B_n=0\,\,(n\neq 1,\,5)$.
Since contributions from $D_{10}$ can be absorbed into $\theta_{01}$
when $p=2$,
we can restrict our consideration to the case
$\theta_{01}=\theta$ and $\theta_{10}=0$.
The quantity
\beqa
  \vev{e^{\,\theta\,D_{01}}}\,=\,
	\frac{\bra{-\frac{B}{g}}\,e^{\,\theta\,D_{01}}\,\ket{\Phi}}{
	\bracket{-\frac{B}{g}}{\Phi}}
  \,=\,\sum_{n=0}^\infty\frac{\theta^n}{n!}\,\vev{D_{01}^n}
\eeqa
is then calculated as follows.
{}First we evaluate
\beqa
  \vev{D_{01}}&=&\gint\ddz \vev{e^{\rphi_0(\zeta)-\rphi_1(\zeta)}}\n\\
  &=&\gint\ddz \exp\left\{\vev{e^{\rphi_0(\zeta)-\rphi_1(\zeta)}-1}_c
	\right\}\n\\
  &=&\gint\ddz \exp\left\{
	\vev{\rphi_0(\zeta)-\rphi_1(\zeta)}
	+\frac{1}{2}\vev{\left(\rphi_0(\zeta)-\rphi_1(\zeta)\right)^2}_c
	+\,\cdots\right\}.
\eeqa
All the leading contributions to the exponent
come from spherical topology, and thus we have
\beqa
  \vev{D_{01}}&=&\gint\ddz\,
  e^{\,(1/g)\,\vev{\rphi_0(\zeta)-\rphi_1(\zeta)}^{(0)}
  \,+\,(1/2)\,\vev{\left(\rphi_0(\zeta)-\rphi_1(\zeta)\right)^2}^{(0)}_c
  \,+\,O(g)}\,.
\eeqa
The first term (denoted $\Gamma_{01}(\zeta)$ in ref.\ \cite{fy})
is calculated by integrating the disk amplitude with respect to
$\zeta$ and found to be \cite{d,fy}
\beqa
  \vev{\rphi_0(\zeta)-\rphi_1(\zeta)}^{(0)}\,=\,
	\frac{16}{15}\,\left(\zeta-\frac{3}{2}\sqrt{t}\right)
	\,\left(\zeta+\sqrt{t}\right)^{3/2}.
\eeqa
On the other hand, to compute the second term, we have to integrate
two-point function (cylinder amplitude) twice.
Explicit form of the cylinder amplitude can be found in ref.\
\cite{mss,dkk-k},
and we obtain\footnote{
{}For the general $(p,q)=(p,p+1)$ case, we have
\beqa
  \vev{\rphi_a(\zeta_1)\,\rphi_b(\zeta_2)}_c^{(0)}&=&
  \log\left[
	\omega^a\left(\zeta_1+\sqrt{\zeta_1^2-t}\right)^{1/p}
 	\,+\,\omega^{-a}\left(\zeta_1-\sqrt{\zeta_1^2-t}\right)^{1/p}
 	\right.\n\\
  &&~~~~~~\left.
	\,-\,\omega^b\left(\zeta_2+\sqrt{\zeta_2^2-t}\right)^{1/p}
	\,-\,\omega^{-b}\left(\zeta_2-\sqrt{\zeta_2^2-t}\right)^{1/p}
  	\right]\n\\
  &&~~-\,\delta_{ab}\,\log\left(\zeta_1-\zeta_2\right)
	\,+\,\left(\delta_{ab}-\frac{1}{p}\right)\,\log R\,,\n
\eeqa
where $\omega=e^{2\pi i/p}$.
This normalization corresponds to
$B_1=t~(>\!0),\,B_{2p+1}=-4p/(p+1)(2p+1)$ and $B_n=0~(n\neq1,\,2p+1)$.}
\beqa
  \vev{\left(\rphi_0(\zeta)-\rphi_1(\zeta)\right)^2}_c^{(0)}\,=\,
  -\,2\,\log \frac{4\,(\zeta+\sqrt{t})}{R},
\eeqa
where $R$ is the infrared cutoff necessary
to define the two-point function.
In the weak coupling limit ($g \rightarrow +\,0$),
we can evaluate the integral by the saddle-point method
at $\zeta=\sqrt{t}/2$
(see \cite{fy} for the detailed investigation on this point),
and find that
\beqa
  \vev{D_{01}}\,=\,\beta\,R\,\sqrt{g}\,t^{-5/8}\,
	e^{-\,4\sqrt{6}\,t^{5/4}/\,5g\,+\,O(g)} .
\eeqa
Here $\beta$ is a definite numerical constant.
On the other hand, $\vev{D_{01}^n}$ vanishes for $n\geq 2$.
In fact, we encounter the following expression
in the process of evaluation:
\beqa
  \vev{D_{01}^n}\,=\,
	\gint\frac{d\zeta_1}{2\pi i}\cdots\frac{d\zeta_n}{2\pi i}\,
	\prod_{i>j}\left(\zeta_i-\zeta_j\right)^2\,
	e^{(1/g)\,\sum_{i=1}^n\,\Gamma_{01}(\zeta_i)\,+\,O(g^0)},
\eeqa
which vanishes in the weak coupling limit,
since in the $p=2$ case we have a single saddle point\footnote{
{}For $p\geq 3$, higher terms $\vev{D_{ab}^n}~(n\geq2)$ could survive
since we have multi saddle points.},
$\zeta_1=\zeta_2=\cdots=\zeta_n=\sqrt{t}/2$.
Therefore, we obtain
\beqa
  \log\left\langle\left.-\frac{B}{g}\,\right|\,\Phi,\theta\,
 	\right\rangle
  \,=\,\log\left\langle\left.-\frac{B}{g}\,\right|\,\Phi\,\right\rangle
  \,+\,\log\,\left[\,1\,+\,
	\theta_{\rm R}\,\sqrt{g}\,t^{-5/8}\,
	e^{-\,4\sqrt{6}\,t^{5/4}/\,5g}\,\right],
\eeqa
where $\theta_{\rm R}\equiv\beta\, R \,\theta$ is the renormalized
theta parameter.
Thus, taking derivatives with respect to the cosmological constant $t$,
for example, we get
\beqa
  u(t,g,\,\theta)&\equiv&\vev{\cO_1\cO_1}_{c,\,\theta} \n\\
  &=&g^2\,\del_t^{\,2}\,
	\log\,\left\langle\left.-\frac{B}{g}\,\right|\,
	\Phi,\theta\,\right\rangle\\
  &=&-\,\frac{\sqrt{t}}{2}\,+\,
	6\,\theta_{\rm R}\,\sqrt{g}\,t^{-1/8}\,
	e^{-\,4\sqrt{6}\,t^{5/4}/\,5g}\,\times\,
	\left(1\,+\,\theta_{\rm R}\,\sqrt{g}\,t^{-5/8}
	e^{-\,4\sqrt{6}\,t^{5/4}/\,5g}\right)^{-2}. \n
\eeqa
The first term represents the asymptotic solution for sphere,
$u_{\rm pert}^{(0)}(t)$,
and the second term exactly reproduces eqs.\ \eq{3} and \eq{4}
with $a=6\,\theta_{\rm R}$.
Recall that we have neglected contributions from higher topologies.

In conclusion, we have shown that we can obtain correct combinatorics
of solitons in our formulation
if we require that soliton background be expressed by a decomposable
state,
namely, the corresponding partition function is a $\tau$ function
of the KP hierarchy.
{}For further study of the dynamics of these solitons,
it would be convenient to construct the string field action in terms of
fermions.
The investigation in this direction is now in progress
and will be reported elsewhere.

\section*{\protect\large{\protect\bf Acknowledgment}}
We would like to thank N.\ Ishibashi, A.\ Ishikawa, K.\ Itoh,
H.\ Itoyama, H.\ Kawai, T.\ Kawano, and M.\ Ninomiya
for useful discussions.
This work is supported in part by the Grant-in-Aid for Scientific
Research from the Ministry of Education, Science and Culture.


\end{document}